\documentstyle[12pt,agums,epsf]{article}

\lefthead{ROTHSCHILD ET AL.}

\righthead{ANTINEUTRINO GEOPHYSICS WITH LIQUID SCINTILLATOR DETECTORS}

\received{September~23,~1997}
\revised{February~10,~1998}
\accepted{February~18,~1998}

\paperid{}

\cpright{PD}{1997}

\authoraddr{F.~P. Calaprice, M.~C. Chen and C.~G. Rothschild,
Physics Department,
Princeton University,
Princeton, New Jersey  08544.  (e-mail: fpc@pupcyc2.princeton.edu;
mchen@post.queensu.ca; caseyr@princeton.edu)}

\slugcomment{Published in {\it Geophysical Research  
Letters}, {\bf 25}, 1083 (1998).}

\begin{document}
\title{Antineutrino geophysics with liquid scintillator detectors}
\author{Casey G.~Rothschild, Mark C.~Chen and Frank P.~ Calaprice}
\affil{Physics Department, Princeton University
Princeton, New Jersey}

%
%

\begin{abstract}
Detecting the antineutrinos emitted by the decay of radioactive elements
in the mantle and crust could provide a direct measurement of the
total abundance of uranium and thorium in the Earth.
In calculating the antineutrino flux at specific sites,
the local geology of the crust and the background from
the world's nuclear power reactors are important considerations.
Employing a global crustal map, with type and thickness data, and using
recent estimates of the uranium and thorium distribution in the Earth,
we calculate the antineutrino event rate for two new neutrino detectors.
We show that spectral features allow terrestrial antineutrino events to be
identified above reactor antineutrino backgrounds and that
the uranium and thorium contributions can be separately determined.
\end{abstract}


\begin{article}
\section{Introduction}
The possibility to study the radiochemical composition of the Earth by
detecting, at the surface, the antineutrinos ($\bar{\nu}_e$) emitted
by the decay of radioactive isotopes within has been suggested before
\cite{Eders,Marx,MarxLux,Avilez,Krauss,Kobayashi}.
Confirming the abundance of certain radioactive elements in the crust
and mantle could establish important geophysical constraints on the
heat generation within the Earth.  Radioactivity is thought to be the
dominant heat source driving mantle convection \cite{DavRich,Sleep}.  It is
also an important factor in determining the surface heat flow in continents
and in understanding the thermal history of the Earth.  Uranium and thorium,
in particular, play an important role due to their relatively large
abundances and long half-lives.  Improved constraints on the present
uranium and thorium abundances in the mantle and crust could yield
better estimates for these internal, geodynamic heat sources.

Detecting the low-energy antineutrinos from radioactivity is a challenge
due to their small interaction cross-section.  At present,
the only practical medium for detecting terrestrial antineutrinos
would be a large-mass liquid scintillator, observing the
inverse $\beta$-decay reaction on the proton.  New scintillator detectors
are being built in Italy and Japan that could inaugurate the
field of antineutrino geophysics.

Various candidates for the target nucleus to be used in an
antineutrino detector have also been discussed;
however, if practical considerations for procuring the huge quantities (tons)
of target material are evaluated, only hydrogen, in the form of water
or organic liquid scintillator (CH$_2$--), emerges as a feasible target.
The reaction:
\begin{equation}
\bar{\nu}_e + \, p \rightarrow e^+ + \, n,
\end{equation}
has a relatively large cross-section.  Its main advantage is
the production of a neutron which captures with a mean lifetime of about
200 $\mu$s, via $n + p \rightarrow d + \gamma$, releasing 2.2~MeV energy
and forming a delayed coincidence with the positron.
This provides a distinctive event signature for an antineutrino interaction.
The threshold for this reaction is 1.8~MeV; the maximum energy of the
antineutrinos produced by natural radioactivity is 3.27~MeV\@.
Consequently, the positron from (1) deposits at most 2.5~MeV
(including both 0.511 MeV annihilation $\gamma$ rays).
At these low energies, liquid scintillator detectors would be
favored over water \v{C}erenkov detectors, due to their
higher light yield.

Of the naturally-occurring radioactive elements in the Earth, only
four $\beta$ decays in the $^{238}$U and $^{232}$Th chains
produce a significant number of antineutrinos with energy greater
than 1.8~MeV\@.  The decay of $^{214}$Bi (U chain) has an
18\% branch \cite{Lederer}
to the ground state, with a $\beta$ endpoint of 3.27~MeV\@.
The $\beta$ decays from $^{234}$Pa (U chain), $^{228}$Ac and
$^{212}$Bi (Th chain) have practically the same maximum decay energy,
2.29 MeV, 2.08 MeV, and 2.25 MeV, respectively.
Consequently, the terrestrial $\bar{\nu}_e$ spectrum above 1.8~MeV
takes on a ``two-component'' form, with its higher-energy component
coming solely from $^{238}$U decay and a lower component with
contributions from $^{238}$U and $^{232}$Th.  This distinctive
spectral shape would also assist in identifying these events in a 
scintillation detector.

\section{Uranium and Thorium Distribution}
It is usually assumed that the Earth's core contains no U and Th
after core/mantle partitioning \cite{Kargel}.  Consequently, the
starting point for determining the distribution of uranium and thorium
in the present crust and mantle is understanding the composition of the
``Bulk Silicate Earth'' (BSE), which is the model representing the
primordial mantle prior to crust formation (equivalent in composition
to the modern mantle plus crust).  BSE compositional estimates have been
compiled \cite{Kargel} and derived using several techniques.
More recently, BSE concentrations of
0.0795 ppm ($\pm 15\%$) for $^{232}$Th and 0.0203~ppm ($\pm 20\%$) for U,
have been suggested \cite{McDSun}, with subjective error estimates
that encompass most of the values listed in \cite{Kargel}.
This amount of thorium in the mantle and crust would produce
9~TW of heat, and uranium 8~TW, which can be compared to the 
total global heat outflow of $\sim$40~TW \cite{Davies,Sclater}.

In the formation of the Earth's crust, the primitive mantle was
depleted of uranium and thorium \cite{ONions}, while the
continental crust, in particular, was highly enriched \cite{Plant}.
Rock samples from the upper continental crust provide direct
isotopic abundance information.  Data from continental heat flow
measurements \cite{Pollack} are also important in deducing the bulk
radiochemical composition.  A recent perspective on
the nature of the continental crust \cite{Rudnick}
includes data from chemical studies of lower crustal rocks and makes
lithological assignments from seismological data
to improve estimates of the composition of the middle and
lower crust.  Combining these sources, the current best estimates for
the average uranium and thorium concentrations in the continental crust
are listed in \callout{Table~1}.

Samples of mid-ocean ridge basalts (MORB) provide the main input
for estimating the bulk uranium and thorium content of oceanic crust.
Concentrations slightly higher than those found in
typical MORB's are representative of the bulk oceanic crust composition
\cite{Jochum,T&M}; values are listed in Table~1.

With the above data, the average U and Th concentrations for
the ``residual'', present-day mantle can be calculated.  We assumed
an average thickness of 40 km for the continental crust \cite{T&M}
and 6 km for the oceanic crust \cite{T&M} and subtracted these enriched
layers from the BSE concentrations.  An interesting problem is
distributing the isotopic depletion of the mantle.
The discontinuity at 670 km depth may act as a barrier to
mass flow \cite{Ringwood}, resulting in U and Th depletion of only
the upper mantle.  Others do not consider this to be the case \cite{Christ}.
Taking the assumption of whole-mantle convection, we selected a uniform
composition for the entire mantle and list average U and Th concentrations
in Table~1.

\section{Results}
The difference in U and Th activity between
continental and oceanic crust is substantial.  Therefore, the
crustal topography near an antineutrino
detector is important to take into account when
calculating the $\bar{\nu}_e$ flux; previous calculations did not
include such a detailed analysis.  We employed crust type and
thickness data in the form of a global crustal map:
(W.~D.~Mooney, G.~Laske, and T.~G.~Masters, CRUST~5.1: A Global Crustal Model
at $5^\circ \times 5^\circ$, submitted to the {\em Journal of Geophysical
Research,} 1997; data accesible from the World Wide Web server for the
U.S. Geological Survey at http://quake.wr.usgs.gov/study/CrustalStructure/).
The data include 16 primary crust types with
about 140 different sub-types.  In our calculations, we lumped the
crust classification into just continental and oceanic and used
the average U and Th concentrations (from Table~1) regardless of
sub-classification, but did use the crust thickness data,
which range from 6--70 km.

We calculated the $\bar{\nu}_e$ flux from U and Th decay
for several sites around the world.  In the $^{238}$U decay chain, there
are 6 $\beta$ decays ($\bar{\nu}_e$ emitting) for each $^{238}$U decay;
the $^{232}$Th decay chain goes through 4 $\beta$ decays per $^{232}$Th decay.
In \callout{Table~2}, these factors were included in the tabulated fluxes.
From this total flux only an average of 0.38 antineutrinos per $^{238}$U decay
(from all branches in the entire decay chain) have energy above the
reaction threshold of 1.804 MeV\@.  For the thorium chain,
an average of 0.15 antineutrinos per $^{232}$Th decay
can make the inverse $\beta$-decay reaction.

Entries for the geographic ``maximum'' and ``minimum'' are also included
in Table~2.  In these two bounding, hypothetical cases, the maximum
refers to the situation where all of the continental crust
(of average thickness 40~km, covering 40\% of the Earth's surface area)
forms a circular cap centered around the $\bar{\nu}_e$ detector.
The minimum refers to the contrasting scenario, with a
$\bar{\nu}_e$ detector in the middle of a grand ocean
(6~km oceanic crust covering 60\% of the Earth).
It is interesting that the terrestrial $\bar{\nu}_e$ flux for a site
in the Himalayas is close to the geographic maximum flux;
this is due to a crust thickness of $\sim$70 km in this region, which
exceeds the average value.

To lowest-order, neglecting the neutron recoil, the cross-section
for $\bar{\nu}_e$ capture on protons is \cite{Vogel}:
\begin{equation}
\sigma(E_\nu) = \frac{2 \, \pi^2 \, \hbar^3}{{m_e}^5 \, c^8 \,f \tau_n} \:
                (E_\nu - \Delta M c^2) \:
                [(E_\nu - \Delta M c^2)^2 - (m_e c^2)^2]^{1/2},
\end{equation}
where $\Delta M$ is the neutron--proton mass difference,
and the $f \tau_n$ values come from neutron $\beta$ decay
\cite{Wilkinson,PDG}.  We integrated this cross-section with the
antineutrino spectra from each energetic U and Th
$\beta$ decay, including Fermi function corrections \cite{behrens},
to estimate event rates in two new liquid scintillator detectors being built.
The Borexino experiment at Gran Sasso will contain 280 tons of
pseudocumene-based scintillator (C$_9$H$_{12}$); in this detector, the
terrestrial antineutrino event rate would be 10 events per year.
For the Kam-LAND experiment in the Kamioka mine, 1,000 tons of
mineral-oil-based scintillator (CH$_2$--) would detect 29 $\bar{\nu}_e$
events from U and Th decay per year.

\section{Discussion}
The terrestrial $\bar{\nu}_e$ event rates are very low; however,
these events have a correlated signature, the positron--neutron
delayed coincidence.  The most problematic background in reactor
antineutrino experiments are fast neutrons (especially
those produced by muon interactions).  To minimize this background requires:
an underground location for cosmic-ray flux reduction; low-radioactivity
in the detector; sufficient external neutron shielding;
and a tight muon veto surrounding the detector.
These conditions might be satisfied in the two new experiments.

A background that cannot be shielded is the $\bar{\nu}_e$ flux from the
world's nuclear reactors \cite{Lagage}.  In Table~2, we also list our
calculated reactor $\bar{\nu}_e$ fluxes.  This calculation employed data
from the International Nuclear Safety Center Database and assumed all reactors
operating at full power (data for all of the world's nuclear reactors are
available from the World Wide Web server for the
International Nuclear Safety Center at http://www.insc.anl.gov).
The scaling of the $\bar{\nu}_e$ flux was
based upon a reactor antineutrino yield of $5 \times 10^{20}$
$\bar{\nu}_e$ per second for 2,800 MW thermal power \cite{Zacek}
and a typical efficiency of 33\% for electric/thermal power.  

The unique two-component shape of the terrestrial antineutrino
energy spectrum makes it possible to identify these events above
the reactor $\bar{\nu}_e$ background.  In \callout{Figure~1},
the energy spectrum for the positrons produced by $\bar{\nu}_e$
capture on protons is displayed,
revealing spectral features at 1.5 and 2.5 MeV\@.
The positron spectrum from reactor antineutrinos has a
well-known shape \cite{Zacek}.  Measuring this spectrum enables the
reactor backgrounds to be subtracted and allows
the higher-energy terrestrial $\bar{\nu}_e$ component from uranium to be
separated from the thorium contribution.

Site selection would be important for future terrestrial
$\bar{\nu}_e$ experiments.  A site in Australia, far from
any nuclear reactors, has a flux $\sim$200$\times$ lower than
the reactor $\bar{\nu}_e$ flux at Kamioka.  An oceanic location
(possibly near Hawaii) enables a $\bar{\nu}_e$ flux measurement
to be predominantly sensitive to radioactivity from the mantle.
It would be interesting to compare the terrestrial antineutrino
data from several experiments around the world.  In doing so, one might
be able to separate the contributions from the crust and mantle.
Though event rates will be low and background suppression
a challenge, two new experiments will be evaluating these
backgrounds and exploring the prospects for antineutrino
geophysics in the near future.

\acknowledgments
We thank Profs.\ F.~A.~Dahlen, K.~S.~Deffeyes and T.~S.~Duffy,
from the Department of Geosciences, Princeton University, for useful
discussions and Prof.\ A.~Suzuki for providing information about
the Kam-LAND experiment.



%
%

   %
   %
   %
{}
\end{article}

\newpage

\begin{figure}
\epsffile{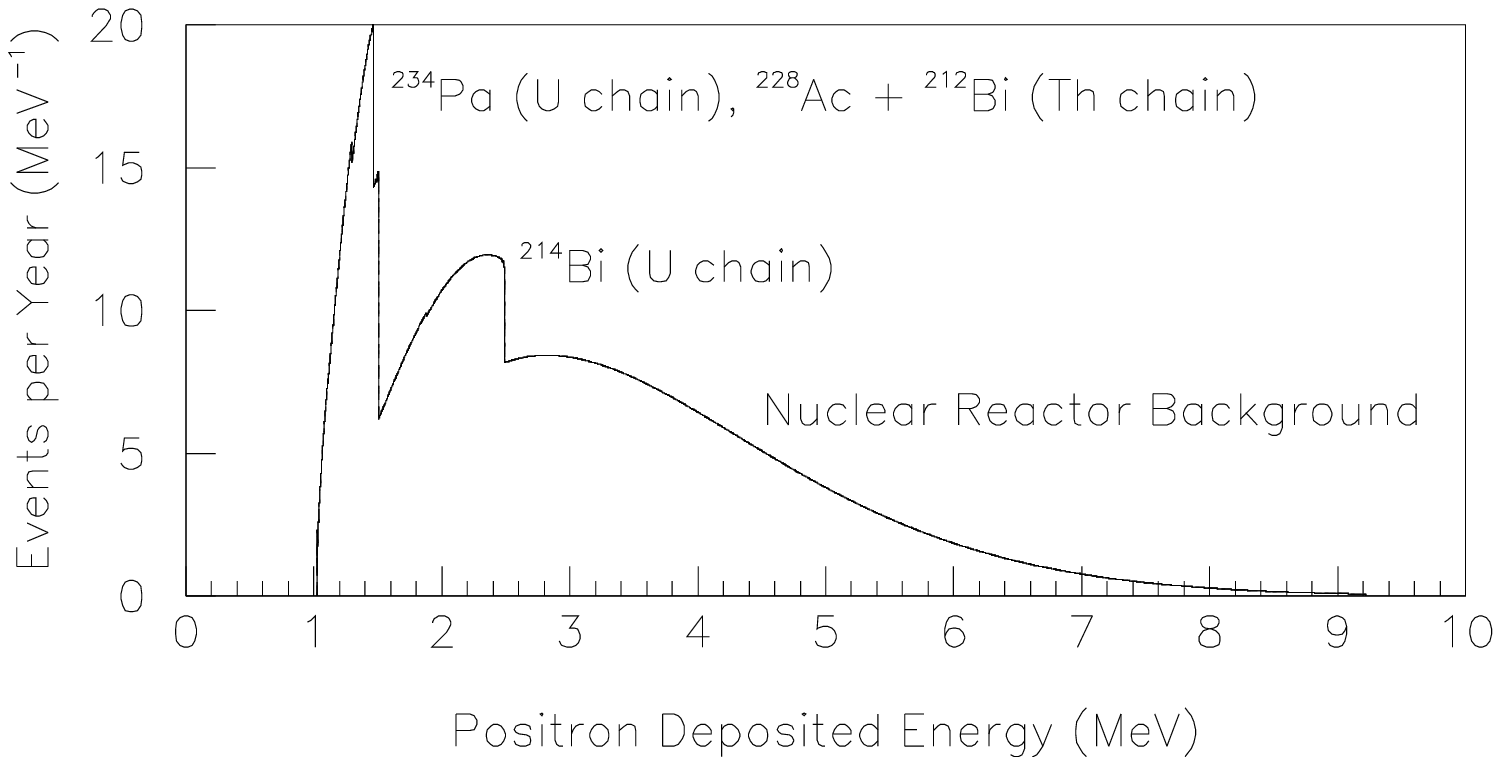}
\caption{Positron energy spectrum from $\bar{\nu}_e$ events
in a 280-ton detector at the Gran Sasso underground laboratory.
The reactor $\bar{\nu}_e$ background rate is 29 events per year,
with only 7.6 events in the same spectral region as the
terrestrial antineutrinos.}
\end{figure}

\newpage

%
%

   %

   %

\begin{table}
\caption{Uranium and thorium abundances in the Earth.}
\begin{tabular}{lll}
\tableline
 				& [$^{238}$U] in ppm & [$^{232}$Th] in ppm \\
\tableline
Bulk Silicate Earth \cite{McDSun}	& 0.0203	  & 0.0795 \\
average continental crust \cite{Rudnick}& 1.4		  & 5.6	   \\
average oceanic crust \cite{T&M}	& 0.10		  & 0.22   \\
present-day ``residual'' mantle		& 0.013		  & 0.052  \\
\end{tabular}
\end{table}

\newpage

\begin{table}
\caption{Calculated $\bar{\nu}_e$ fluxes [$\times$ 10$^6$ cm$^{-2}$ s$^{-1}$]
for sites around the world.}
\begin{tabular}{llllllll}
\tableline
 & & \multicolumn{2}{c}{Uranium} & \multicolumn{2}{c}{Thorium} & 
     Total & Reactor \\
Site & Location & crust & mantle & crust & mantle &
     (U + Th)   & background \\
\tableline
Gran Sasso Lab (Italy) 	& 42$^\circ$N 14$^\circ$E	&
2.5  & 1.2 & 2.2  & 1.0 & 6.9 & 0.65  \\
Kamioka Mine (Japan) 	& 36$^\circ$N 137$^\circ$E	&
0.82 & 1.2 & 0.69 & 1.0 & 3.7 & 3.7   \\
Sudbury (Canada)	& 47$^\circ$N 81$^\circ$W	&
3.2  & 1.2 & 2.8  & 1.0 & 8.2 & 1.3   \\
Central Australia 	& 25$^\circ$S 133$^\circ$E	&
2.7  & 1.2 & 2.4  & 1.0 & 7.3 & 0.016 \\
Himalayas (Tibet)	& 33$^\circ$N 85$^\circ$E	&
3.5  & 1.2 & 3.1  & 1.0 & 8.8 & 0.054 \\
Pacific Ocean (Hawaii)	& 20$^\circ$N 156$^\circ$W	&
0.30 & 1.2 & 0.23 & 1.0 & 2.7 & 0.027 \\
Geographic Maximum	&				&
3.5  & 1.2 & 3.1  & 1.0 & 8.8 &       \\
Geographic Minimum	&				&
0.21 & 1.2 & 0.16 & 1.0 & 2.6 &       \\
\end{tabular}
\end{table}

\clearpage

\begin{thebibliography}{}
%
\bibitem[{\it Avilez et al.,} 1981]{Avilez}
\reference
Avilez, C., G.~Marx and B.~Fuentes, Earth as a source of antineutrinos, 
{\em Phys.\ Rev.\ D,} {\bf 23}, 1116--1117, 1981.
%
\bibitem[{\it Behrens and J\"anecke,} 1969]{behrens}
\reference
Behrens, H., and J.~J\"anecke, 
{\em Numerical Tables for Beta-Decay and Electron Capture,} 316 pp.,
Springer-Verlag, Berlin, 1969.
%
\bibitem[{\it Christensen,} 1989]{Christ}
\reference
Christensen, U.~R., Models of mantle convection: One or several layers,
{\em Phil.\ Trans.\ R.\ Soc.\ Lond.,} {\bf 328}, 417--424, 1989.
%
\bibitem[{\it Davies,} 1980]{Davies}
\reference
Davies, G.~F., Thermal histories of convective Earth models and constraints on
radiogenic heat production in the Earth,
{\em J.\ Geophys.\ Res.,} {\bf 85}, 2517--2530, 1980.
%
\bibitem[{\it Davies and Richards,} 1992]{DavRich}
\reference
Davies, G.~F., and M.~A.~Richards,
Mantle convection,
{\em J.\ Geol.,} {\bf 100}, 151--206, 1992.
%
\bibitem[{\it Eders,} 1966]{Eders}
\reference
Eders, G., Terrestrial neutrinos,
{\em Nucl.\ Phys.,} {\bf 78}, 657--662, 1966.
%
\bibitem[{\it Jochum et al.,} 1983]{Jochum}
\reference
Jochum, K.~P., A.~W.~Hoffman, E.~Ito, H.~M.~Seufert, and W.~M.~White,
K, U and Th in mid-ocean ridge basalt glasses and heat production, K/U and K/Rb
in the mantle,
{\em Nature,} {\bf 306}, 431--436, 1983.
%
\bibitem[{\it Kargel and Lewis,} 1993]{Kargel}
\reference
Kargel, J.~S., and J.~S.~Lewis, The composition and early evolution of Earth,
{\em Icarus,} {\bf 105}, 1--25, 1993.
%
\bibitem[{\it Krauss et al.,} 1984]{Krauss}
\reference
Krauss, L.~M., S.~L.~Glashow, and D.~N.~Schramm, Antineutrino astronomy and
geophysics, 
{\em Nature,} {\bf 310}, 191--198, 1984.
%
\bibitem[{\it Kobayashi and Fukao,} 1991]{Kobayashi}
\reference
Kobayashi, M., and Y.~Fukao, The Earth as an antineutrino star,
{\em Geophys.\ Res.\ Lett.,} {\bf 18}, 633--636, 1991.
%
\bibitem[{\it Lagage,} 1985]{Lagage}
\reference
Lagage, P.~O., Nuclear power stations as a background source for antineutrino
astronomy,
{\em Nature,} {\bf 316}, 420--421, 1985.
%
\bibitem[{\it Lederer and Shirley,} 1978]{Lederer}
\reference
Lederer, C.~M., and V.~S.~Shirley (Eds.),
{\em Table of Isotopes, Seventh Edition,} 1523 pp., John Wiley, New York, 1978.
%
\bibitem[{\it Marx,} 1969]{Marx}
\reference
Marx, G., Geophysics by neutrinos,
{\em Czech.\ J.\ Phys.\ B,} {\bf 19}, 1471--1479, 1969.
%
\bibitem[{\it Marx and Lux,} 1970]{MarxLux}
\reference
Marx, G., and I.~Lux, Hunting for soft antineutrinos,
{\em Acta Phys.\ Acad.\ Sci.\ Hung.,} {\bf 28}, 63--70, 1970.
%
\bibitem[{\it McDonough and Sun,} 1995]{McDSun}
\reference
McDonough, W.~F., and S.-s.~Sun, The composition of the Earth,
{\em Chem.\ Geol.,} {\bf 120}, 223--253, 1995.
%
\bibitem[{\it O'Nions and McKenzie,} 1993]{ONions}
\reference
O'Nions, R.~K., and D.~McKenzie, Estimates of mantle thorium/uranium ratios from
Th, U and Pb isotope abundances in basaltic melts, 
{\em Phil.\ Trans.\ A,} {\bf 342}, 65--77, 1993.
%
\bibitem[{\it Particle Data Group,} 1996]{PDG}
\reference
Particle Data Group, Review of Particle Physics,
{\em Phys.\ Rev.\ D,} {\bf 54}, 1--720, 1996.
%
\bibitem[{\it Plant and Saunders,} 1996]{Plant}
\reference
Plant, J.~A., and A.~D.~Saunders, The radioactive Earth, 
{\em Radiation Protection Dosimetry,} {\bf 68}, 25--36, 1996.
%
\bibitem[{\it Pollack and Chapman,} 1977]{Pollack}
\reference
Pollack, H.~N., and D.~S.~Chapman, On the regional variation of heat flow,
geotherms, and lithospheric thickness,
{\em Tectonophysics,} {\bf 38}, 279--296, 1977.
%
\bibitem[{\it Ringwood and Irifune,} 1988]{Ringwood}
\reference
Ringwood, A.~E., and  T.~Irifune, Nature of the 650 km seismic discontinuity:
Implications for mantle dynamics and differentiation,
{\em Nature,} {\bf 331}, 131--137, 1988.
%
\bibitem[{\it Rudnick and Fountain,} 1995]{Rudnick}
\reference
Rudnick, R.~L., and D.~M.~Fountain,
Nature and composition of the continental crust: A lower crustal perspective,
{\em Rev.\ Geophys.,} {\bf 33}, 267--309, 1995.
%
\bibitem[{\it Sclater et al.,} 1980]{Sclater}
\reference
Sclater, J.~G., C.~Jaupart, and D.~Galson,
The heat flow through oceanic and continental crust
and the heat loss of the Earth,
{\em Rev.\ Geophys.\ Space Sci.,} {\bf 18}, 269--311, 1980.
%
\bibitem[{\it Sleep,} 1990]{Sleep}
\reference
Sleep, N.~H., Hotspots and mantle plumes: Some phenomenology,
{\em J.\ Geophys.\ Res.,} {\bf 95}, 6715--6736, 1990.
%
\bibitem[{\it Taylor and McLennan,} 1985]{T&M}
\reference
Taylor, S.~R., and S.~M.~McLennan, 
{\em The Continental Crust: its Composition and Evolution,} 312 pp.,
Blackwell Scientific, Oxford, 1985. 
%
\bibitem[{\it Vogel,} 1984]{Vogel}
\reference
Vogel, P., Analysis of the antineutrino capture on protons,
{\em Phys.\ Rev. D,} {\bf 29}, 1918--1922, 1984.
%
\bibitem[{\it Wilkinson,} 1982]{Wilkinson}
\reference
Wilkinson, D.~H., Analysis of neutron $\beta$-decay,
{\em Nucl.\ Phys.\ A,} {\bf 377}, 474--504, 1982.
%
\bibitem[{\it Zacek et al.,} 1986]{Zacek}
\reference
Zacek, G.\ {\em et al.,} Neutrino-oscillation experiments at the G\"osgen
nuclear power reactor,
{\em Phys.\ Rev.\ D,} {\bf 34}, 2621--2636, 1986.
%
\end{thebibliography}
\end{document}